\documentclass[]{revtex4}
\usepackage{graphicx}
\usepackage{fancyhdr}
\begin{document}

\title{ {\bf Wilson Coefficients in the Operator Product Expansion of Scalar Currents at Finite Temperature    }}
\author{El\c{s}en Veli Veliev *,  Takhmassib M. Aliev \\ ** Physics Department, Kocaeli University, Umuttepe Yerle\c{s}kesi, 41380 Izmit, Turkey \\ Physics Department, Middle
East Technical University, 06531 Ankara, Turkey\\} \fancyhead[C]{\it
{International Conference on Hadron Physics, TROIA'07, 30 Aug - 3
Sep 2007, Canakkale, Turkey}}
\begin{abstract}
 In this paper, we have investigated operator product expansion for thermal
 correlation function of the two scalar currents. Due to breakdown of
 Lorentz invariance at finite temperature, more operators of the same
 dimension appear in the operator product expansion than those at zero temperature.
 We calculated Wilson coefficients in the short distance expansion and obtain
 operator product expansion for thermal   correlation function in terms of
  quark condensate $\langle \overline{\psi}\psi \rangle$, gluon condensate
  $\langle G^{a}_{\mu\nu}G^{a\mu\nu}\rangle$, quark energy density $\langle u\Theta^{f}u\rangle$
  and  gluon energy density $\langle u\Theta^{g}u\rangle$.
 \end{abstract}

\maketitle

\thispagestyle{fancy}


\section{Introduction}

The Shifman, Vainshtein and Zakharov (SVZ) sum rules, proposed about
three decades ago \cite{1}, is one of the powerful method for
investigating the properties of hadrons. This method has been
extensively used as an efficient tool to study properties of
resonances, decay form factors and so on \cite{2}. In this approach,
hadrons are represented by their interpolating quark currents taken
at large virtualities and the correlation function of these currents
is investigated in the framework of operator product expansion
(OPE).

The SVZ sum rules method is extended to the finite temperature in
the paper \cite{3}. In extending these sum rules to finite
temperature, two sources of complications arise. One is the
interaction of the current with the particles of the medium. The
other complication is the breakdown of Lorentz invariance by the
choice of the reference frame \cite{4, 5}. Taking into account both
complications, OPE of vector currents at finite temperature first
investigated in \cite{6} and application of these results to the
temperature dependence of the  $\rho$-meson parameters are done in
\cite{7, 8}. Also, modifications of meson parameters due to nuclear
medium are widely discussed in the literature \cite{9, 10}.

In this paper we studied OPE for thermal correlation function,
necessary for the phenomenological investigation of scalar meson
parameters. We calculated Wilson coefficients in the OPE of scalar
currents at finite temperature.
\section{Thermal Correlation Function in Short Distance Expansion }
The starting point for SVZ sum rules is the OPE [1] and it gives a
general  form  of  the  considered  quantities  in terms of
operators. We begin by considering the thermal correlation function,
\begin{equation}\label{eqn1}
T(q)=i \int d^{4}x e^{iq\cdot x} \langle T(J(x)J(0))\rangle, \\
\end{equation}
where $J(x)=:\overline{q}(x)q(x):$ is the scalar current and $T$
indicates  the time ordered product. The thermal average of an
operator   is defined as follows \cite{11}
\begin{equation}\label{eqn2}
\langle A\rangle=\frac{tr e^{-\beta H}A}{tr e^{-\beta H}}, \\
\end{equation}
where $H$ is the QCD Hamiltonian and $\beta=1/T$ stands for the
inverse of the temperature. Traces are carried out over any complete
set of states. Using Wick's theorem and making some transformations,
we get
\begin{equation}\label{eqn3}
T(J(x)J(0))=trS(x,0)S(0,x)-\frac{1}{2 \pi^2 x^4} \left[ m x^2
:\overline{\psi}\psi:+2(\not\!{x})_{ab}x^{\mu}:\overline{\psi}_{a}i
D_{\mu}\overline{\psi}_{b}:\right] ,
\end{equation}
where $a$ and $b$ are spinor indices, and $D_{\mu}$ is covariant
derivative. The fundamental assumption of Wilson expansion is that
the product of operators at different points can be expanded as the
sum of local operators with momentum dependent coefficients in the
form:
\begin{equation}\label{eqn4}
T(q)=\sum C_{n}(q^2)\langle O_{n}\rangle , \\
\end{equation}
where $C_{n}(q^2)$ are called Wilson coefficients and $O_{n}$ are a
set of local operators. In this expansion, the operators are ordered
according to their dimension $d$. The lowest dimension operator with
$d=0$ is the unit operator associated with the perturbative
contribution. In the vacuum sum rules low dimension operators
composed of quark and gluon fields are quark condensate $\langle
\overline{\psi}\psi \rangle$ and gluon condensate $\langle
G^{a}_{\mu\nu}G^{a\mu\nu}\rangle$. At finite temperature Lorentz
invariance is broken by the choice of a preferred frame of reference
and new operators appear in the Wilson expansion. To restore Lorentz
invariance in thermal field theory, four-vector velocity of the
medium $u^{\mu}$ is introduced. Using four-vector velocity and
quark/gluon fields, we can construct a new set of low dimension
operators $\langle u\Theta^{f}u\rangle$ and $\langle u\Theta^{g}u
\rangle$ with dimension $d=4$. So, we can write thermal correlation
function in terms of operators up to dimension four:
\begin{equation}\label{eqn5}
T(q)=C_{1} I + C_{2}\langle \overline{\psi}\psi\rangle +
C_{3}\langle G^{a}_{\mu\nu}
G^{a\mu\nu}\rangle + C_{4}\langle u\Theta^{f}u\rangle + C_{5}\langle u\Theta^{g}u\rangle   . \\
\end{equation}
In our calculations, we used the following expansion of quark fields
$\psi(x)$
\begin{equation}\label{eqn6}
\psi(x)=\psi(0)+x^{\mu}D_{\mu}\psi(0)+...  , \\
\end{equation}
which takes place in Fock-Schwinger gauge
$x^{\mu}A_{\mu}^{a}(x)=0$. In order to calculate Wilson coefficients
in \cite{4}, we use massless quark field propagator in the
background gauge field \cite{6}
\begin{equation}\label{eqn7}
S(x,0)=\frac{1}{4 \pi^2}\left[-\frac{2 \not\!{x}}{x^4}-\frac{i g}{4
x^2}\gamma^{\mu}\not\!{x}
 \gamma^{\nu}G_{\mu\nu}+\frac{B}{48}(a \not\!{x}+b\not\!{u})\right] , \\
\end{equation}
where $G_{\mu\nu}=\frac{1}{2}\lambda^{a}G_{\mu\nu}^{a}$ is the gluon
strength tensor,  coefficients $a$  and $b$  have the following
forms
\begin{equation}\label{eqn8}
a=1-\frac{4}{x^2}(u\cdot x)^2-2 \ln(-4x^2\mu^2), \\
\end{equation}
\begin{equation}\label{eqn9}
b=8(u \cdot x) \ln(-4x^2\mu^2). \\
\end{equation}
In our calculations we get expressions of kind
$T_{\alpha\beta\lambda\sigma}=\langle
trG_{\alpha\beta}G_{\lambda\sigma}\rangle$, where trace is carry out
over color matrices. $T_{\alpha\beta\lambda\sigma}$ is four rank
tensor and must be expressed in terms of the metric tensor
$g_{\alpha\beta}$ and four-vector velocities $u_{\alpha}$. The
Lorentz covariance at finite temperature allows us to write general
structure of this tensor in the following form:
\begin{eqnarray}\label{eqn10}
T_{\alpha\beta\lambda\sigma}&=&B_{1}g_{\alpha\beta}g_{\lambda\sigma}+
B_{2}g_{\alpha\lambda}g_{\beta\sigma}+B_{3}g_{\alpha\sigma}g_{\beta\lambda}+
B_{4}g_{\alpha\beta}u_{\lambda}u_{\sigma}+B_{5}g_{\alpha\lambda}
u_{\beta}u_{\sigma}+ B_{6}g_{\alpha\sigma}u_{\beta}u_{\lambda}
\nonumber \\
&+& B_{7}g_{\beta\lambda}u_{\alpha}u_{\sigma} +
B_{8}g_{\beta\sigma}u_{\alpha}u_{\lambda}+
B_{9}g_{\lambda\sigma}u_{\alpha}u_{\beta}+
B_{10}u_{\alpha}u_{\beta}u_{\lambda}u_{\sigma}
\end{eqnarray}
where $B_{i}$ are unknown scalar coefficients. Note that
$T_{\alpha\beta\lambda\sigma}$ is antisymmetric under the
interchange of indices $\alpha$ and $\beta$, as well as indices
$\lambda$ and $\sigma$:
$T_{\alpha\beta\lambda\sigma}=-T_{\beta\alpha\lambda\sigma}$ and
$T_{\alpha\beta\lambda\sigma}=-T_{\alpha\beta\sigma\lambda}$. Using
these properties we obtain that $B_1=B_4=B_9=B_{10}=0$,
$B_6=B_7=-B_5=-B_8$ and $B_2=-B_3$. Therefore
$T_{\alpha\beta\lambda\sigma}=<trG_{\alpha\beta}G_{\lambda\sigma}>$
is expressed in terms of two scalar functions of $B_{2}\equiv A$ and
$B_{6}\equiv B$. Contracting indices on both sides and using gluonic
part of energy momentum tensor $\Theta^{g}_{\mu\nu}$, these
coefficients can be expressed with gluon condensate and gluon energy
densities as follows
\begin{equation}\label{eqn11}
A=\frac{1}{24} \langle
G^{a}_{\mu\nu}G^{a\mu\nu}\rangle+\frac{1}{6}\langle u^{\mu}
\Theta^{g}_{\mu\nu}u^{\nu}\rangle , \\
\end{equation}
\begin{equation}\label{eqn12}
B=\frac{1}{3}\langle u^{\mu}
\Theta^{g}_{\mu\nu}u^{\nu}\rangle . \\
\end{equation}
Using these expressions for coefficients and making some
transformations, we obtain the gluonic part of correlation function
\begin{eqnarray}\label{eqn13}
\langle tr S(x,0)S(0,x) \rangle&=&-\frac{1}{16 \pi^4} \Bigg
[\frac{16}{x^6}-\frac{aB}{3 x^2} -\frac{g^2}{6 x^2}\langle
u\Theta^{g}u \rangle+\frac{g^2}{8 x^2} \langle
G^{a}_{\mu\nu}G^{a\mu\nu}\rangle \nonumber \\ &-& \frac{(u \cdot x)b
B}{3 x^4}+\frac{2 g^2}{3 x^4}(u \cdot x)^2\langle
u\Theta^{g}u\rangle \Bigg ]
\end{eqnarray}
Let us consider second term in eq.(\ref{eqn3}) for zero chemical
potential case. Expanding the quark bilinear matrix in terms of
complete set of Dirac matrices and taking into account parity
conservation, the most general finite temperature decomposition of
two operators in spinor space becomes \cite{6}
\begin{equation}\label{eqn14}
\langle \overline{\psi}_{a}\psi_{b}\rangle=D_1 \delta_{ab}+D_2 (\not\!{u})_{ba} , \\
\end{equation}
\begin{equation}\label{eqn15}
\langle \overline{\psi}_{a}i D_{\mu}\psi_{b}\rangle=E_1
(\gamma_{\mu})_{ba}+E_2
u_{\mu}(\not\!{u})_{ba}+E_3u_{\mu}\delta_{ba}+E_4(\not\!{u}
\gamma_{\mu})_{ba}  \\
\end{equation}
Here, $D_i$ and $E_k$ are scalar coefficients. By contracting
indices on both sides and using fermionic part of energy momentum
tensor $\Theta^{f}_{\mu\nu}$, we get
\begin{equation}\label{eqn16}
E_1=\frac{m}{16}\langle
\overline{\psi}\psi\rangle-\frac{1}{12}\langle
u^{\mu}\Theta^{f}_{\mu\nu}
u^{\nu}\rangle  , \\
\end{equation}
\begin{equation}\label{eqn17}
E_2=\frac{1}{3}\langle u^{\mu}\Theta^{f}_{\mu\nu}
u^{\nu}\rangle  , \\
\end{equation}
Substituting these equalities in eq.(\ref{eqn3}), the contributions
of operators up to dimension four to correlation function in the
coordinate space can be written as
\begin{eqnarray}\label{eqn18}
\langle T( J(x)J(0)) \rangle&=&-\frac{1}{ 2\pi^2x^4}\left[\frac{
3}{2}mx^2 \langle \overline{\psi}\psi\rangle -\frac{2}{3}x^2 \langle
u\Theta^{f}u\rangle+\frac{8}{3}(u\cdot x)^2 \langle
u\Theta^{f}u\rangle\right]
\nonumber \\
&-& \frac{1}{16 \pi^4}\Bigg[\frac{16}{x^6}-\frac{a
B}{3x^2}-\frac{(u\cdot x)b B}{3x^4} +\frac{g^2}{8x^2}\langle
G^{a}_{\mu\nu}G^{a\mu\nu}\rangle
\nonumber \\
&-&\frac{g^2}{6x^2}\langle u\Theta^{g}u\rangle
+\frac{2g^2}{3x^4}(u\cdot x)^2\langle u\Theta^{g}u\rangle \Bigg] .
\end{eqnarray}
Applying the Fourier transformation to eq.(\ref{eqn18}) correlation
function may be written in the momentum space. As can be seen in
Fourier transformation, we meet different kind of integrals such as
\begin{equation}\label{eqn19}
I=\int d^4 x \frac{\ln(-x^2)}{x^2}e^{iq\cdot x}. \\
\end{equation}
Up to now, we evaluated expressions in the Minkowski space. After
that, in order to calculate integrals we proceed to the Euclidean
space \cite{12}. Using the well known identity
\begin{equation}\label{eqn20}
\ln x^2=-\lim_{\lambda\rightarrow 0}\frac{d}{d\lambda}
\frac{1}{(x^2)^{\lambda}} , \\
\end{equation}
the integral in eq.(\ref{eqn19}) can be transformed  to the
following form
\begin{equation}\label{eqn21}
I=-i \, \, \lim_{\lambda\rightarrow 0}\frac{d}{d\lambda}\int dx
\frac{e^{-iQ\cdot x}}{(x^2)^{1+\lambda}} , \\
\end{equation}
where $Q$ is Euclidean momentum and $Q^2=-q^2$. Using the standard
Fourier transformation and expansion formula
$\Gamma(1+\varepsilon)=1-\gamma \varepsilon+O(\varepsilon^2)$, we
get
\begin{equation}\label{eqn22}
I=-i\pi^2 \left(\frac{8 \gamma}{Q^2}+\frac{4}{Q^2}\ln\frac{Q^2}{4}\right) , \\
\end{equation}
where $\gamma$  is Euler constant. After calculating the other
integrals thermal correlation function in the momentum space in
short distance expansion can be written as
\begin{eqnarray}\label{eqn23}
T(Q)&=&\frac{1}{
8\pi^2}Q^2\left(\gamma-\ln\frac{4\pi}{Q^2}\right)+\frac{3}{Q^2}m
\langle \overline{\psi}\psi\rangle+\frac{4}{3} \left(\frac{4 (u\cdot
Q)^2}{Q^4}+\frac{1}{Q^2}\right)\langle u\Theta^{f}u\rangle
\nonumber \\
&+&\frac{g^2}{32 \pi^2 Q^2}\langle
G^{a}_{\mu\nu}G^{a\mu\nu}\rangle-\frac{g^2}{72
\pi^2}\Bigg[(5+8\gamma)\frac{1}{Q^2}-(1+8\gamma)\left(\frac{4(u\cdot
Q)^2}{Q^4}+\frac{2}{Q^2}\right)  \
\nonumber \\
&-& 4 \ln\left(\frac{Q^2}{16 \mu^2}\right)\left(\frac{4 (u\cdot
Q)^2}{Q^4}+\frac{1}{Q^2}\right)+16 \frac{(u\cdot
Q)^2}{Q^4}\Bigg]\langle u\Theta^{g}u\rangle
\end{eqnarray}
which reproduces well-known zero temperature result  in the
$T\rightarrow 0$ limit \cite{13}. At the end, due to breakdown of
Lorentz invariance at finite temperature, two new operators appear
in the Wilson expansion, in addition to the two old (Lorentz
invariant) ones, already existing in the vacuum sum rules. Therefore
OPE for thermal correlator is expressed with four temperature
dependent quantities, quark condensate $\langle
\overline{\psi}\psi\rangle$, gluon condensate $\langle
G^{a}_{\mu\nu}G^{a\mu\nu}\rangle$, quark energy density $\langle
u\Theta^{f}u \rangle$ and gluon energy density $\langle u\Theta^{g}u
\rangle $. The thermal average of quark condensate $\langle
\overline{\psi}\psi\rangle$ is known from chiral perturbation theory
\cite{14, 15}
\begin{equation}\label{eqn24}
\langle\overline{\psi}\psi\rangle=\langle
0|\overline{\psi}\psi|0\rangle\left[1-\frac{n_{f}^{2}-1}{n_{f}}
\frac{T^2}{12 F^2}+O(T^4)\right]\\
\end{equation}
where $n_{f}$  is number of quark flavors and $F=0.088GeV$.

The relationship between the trace of energy momentum tensor and the
gluon condensate has been studied at finite temperature in paper
\cite{16}
\begin{equation}\label{eqn25}
\frac{g^2}{4 \pi^2}\left(\langle
G^{a}_{\mu\nu}G^{a\mu\nu}\rangle-\langle 0|G^{a}_{\mu\nu}
G^{a\mu\nu}|0\rangle\right)=-\frac{8}{9}\left(\langle
\Theta^{\mu}_{\mu}\rangle+\sum_{f}m_{f}
\left(\langle\overline{\psi}\psi\rangle-\langle0|\overline{\psi}\psi|0\rangle\right)\right).\\
\end{equation}
For two massless quarks in the low temperature chiral perturbation
expansion the trace of the total energy momentum tensor
$\Theta=n_{f}\Theta^{f}+\Theta^{g}$ has following form \cite{17}
\begin{equation}\label{eqn26}
\langle\Theta^{\mu}_{\mu}\rangle=\frac{\pi^2}{270}\frac{T^8}{F_{\pi}^{4}}
\ln\frac{\Lambda_{p}}{T}+O\left(T^{10}\right),\\
\end{equation}
where the pion decay constant has the value of $F_{\pi}=0.093GeV$
and the logarithmic scale factor is $\Lambda_{p}=0.275GeV$. The
obtained results allow us to investigate the temperature dependence
of the $\sigma$ and $a_0$ mesons parameters. The investigations of
hadronic parameters at finite temperature are very important for the
interpretation of heavy ion collision results and for understanding
the phenomenological and theoretical aspects of thermal QCD.
\section{Acknowledgement}
This work is supported by the Scientific and Technological Research
Council of Turkey (TUBITAK), research project no.105T131, and the
Research Fund of Kocaeli University under grant no. 2004/4.

\end{document}